\documentclass[aps,prb,twocolumn,groupedaddress,showpacs]{revtex4-1}

\usepackage{graphicx}
\usepackage[geometry]{ifsym}
 \usepackage{epsfig}
 \usepackage{color}
 \usepackage{soul}
 \usepackage{amsmath}
 \usepackage{amssymb}
\usepackage{amsfonts}
\usepackage{listings}
\begin{document}

\title{A computational approach to a doped antiferromagnet: correlations between two spin-polarons in the lightly doped CuO$_2$ plane}

\author{Bayo Lau$^{*,1}$, Mona Berciu$^{1,2}$, and George A. Sawatzky$^{1,2}$}
\affiliation{$^*$currently at Columbia University, contact at bayo@phys.columbia.edu}
\affiliation{$^1$Department of Physics and Astronomy, University of British Columbia, Vancouver, British Columbia, V6T 1Z1}
\affiliation{$^2$Quantum Matter Institute, University of British Columbia, Vancouver, British Columbia, V6T 1Z4}

\date{\today}

\begin{abstract}
We extend the methods recently introduced in Phys. Rev. Lett. 106
036401 (2011) to investigate correlations between two spin-polarons in a
quasi-two-dimensional CuO$_2$ layer.  The low-energy wavefunctions for
two doped holes introduced in a half-filled CuO$_2$ plane with 32
copper and 64 oxygen sites are calculated explicitly using an efficient
yet accurate truncation scheme to model the antiferromagnet
background. The energetics and wavefucntions show that the
charges form three-spin polarons and the spin is carried by a disturbance around the three-spin polaron core. The low-energy band results from the competition between the kinetic energy and a local attractive potential
which favors $d_{x^2-y^2}$ states. Lastly, we point out features
that are expected to be robust for larger systems.
\end{abstract}

\pacs{71.10.Fd,75.10.Jm,71.38.-k,74.72.-h}

\maketitle

\section{Introduction}
Spin-$\frac{1}{2}$ antiferromagnetism is of great importance in the
quantum description of many exotic materials. Setting as a reference
point the exhaustively studied nearest-neighbor Heisenberg
antiferromagnet (AFM) on the two-dimensional square lattice,\cite{Mans} major
developments include studies of frustration due to lattice geometry
and to long(er)-range coupling, as well as the disturbance due to
additional fermionic charge carriers. Investigations of the latter
problem have been strongly driven by the need to understand the
doping-controlled evolution of cuprate layers. This is also the focus
of this article.

Since their discovery in 1986,\cite{bednorzNmuller} cuprates have been
classified as high-$T_c$ superconductors. Their main challenge to
condensed matter physics is to understand the basic mechanism
leading to high temperature superconductivity, but equally important
is the need to understand
the many anomalous properties when these materials are tuned away from
the superconducting phase. The class of hole-doped cuprates, which
allow electron removal from the parent compound, is of particular
interest because of the clear separation of the pseudogap regime, in
addition to the antiferromagnetism, superconductivity, Fermi liquid
and non-Fermi liquid phases which occupy different regions of the
phase diagram. The connections between these different phases have not yet
been fully elucidated despite many
proposals.\cite{dump0,dump1,dump2,dump3,dump4,dump5,dump6,varma,phonon} In
fact, one of the few widely accepted ideas in this field is that to
find the pairing mechanism will require
understanding the various non-superconducting phases first.

It is also widely believed that a complete description of the lightly
hole-doped spin-${1\over 2}$ 2D antiferromagnet (AFM) with full
quantum fluctuations could provide clues for understanding the origin
of the non-Fermi-liquid behavior and the superconducting ground state
observed at higher doping. Consideration of more exotic
scenarios\cite{varma,phonon} are exciting developments; however, a
detailed modeling of the few holes+AFM problem is a crucial first step
needed to appreciate the significance and importance of such additions. This problem
is difficult because of the complicated nature of the 2D AFM
background, whose quantum fluctuations in the presence of multiple
doping holes were never fully captured for a large CuO$_2$ lattice.

Such a theoretical or computational description is challenging because
of the strongly-correlated many-body nature of the system. Microscopic
hole-AFM interactions have been studied in models with one,
\cite{zrs,russianpolaron,tjrev0,tjrev1,tjrev2,tjed,ctmc,2hed} two,
\cite{zaanen2band0,zaanen2band1,zaanen2band2,zaanen2band3,zaanen2band4,macridin2band0,macridin2band1,macridin2band2}
three, \cite{threeband0,threeband1,thomale0,thomale1} or more
\cite{aharony,beyond0,beyond1,beyond2,beyond3,beyond4,beyond5}
bands. While exact analytical solutions seem to be out of reach,
numerical studies are also carried out with various compromises, such
as the use of small clusters, variational approaches, and/or modeling
of the AFM state as a classical N\'eel state plus
spin-waves.\cite{numrev0,numrev1,numrev2} Due to these limitations,
there are uncertainties regarding the minimal microscopic model; after
all, it took decades to gain a satisfactory understanding of even the
simplest one-band t-t'-J model.\cite{tjrev0,tjrev1,tjrev2} While
certain aspects
\cite{details0,details1,stripe,xray,davis,greven0,greven1,fauque,peets} of X-ray spectroscopy, Electron Energy Loss
Spectroscopy, Scanning Tunneling Microscopy, recent neutron
scattering and x-ray absorption measurements cannot be described by
one band models, the significance of omitting other bands cannot be
quantified without access to solutions of more detailed
models.

Cuprates exhibit charge-transfer band-gap behavior with mobile holes
located mainly on the anion $2p$ orbitals and unpaired electrons on
cation $3d$ orbitals.\cite{zsa}  One-band models use superexchange
\cite{superx} and Zhang-Rice singlets (ZRS)\cite{zrs} to reduce the
$(N-n)$-electron problem to one of $n$ holes in an AFM background, in
which both spin and charge degrees of
  freedom are assumed to be hosted in the same single band. To reach
agreement with experiments, such models must be
tweaked at least by adding longer-range
hopping\cite{tp0,tp1,macridin2band0,macridin2band1,macridin2band2} and
possibly by coupling to phonons\cite{shen0,shen1,phonon}.

We have
recently shown that even for the one-hole case, distinguishing cation
and anion sites leads to significantly different wavefunctions
compared to those of such single band models\cite{spinpolaron}.
In particular, we found that the low-energy quasiparticle band
is a result of the crossing between the bands of spin-$\frac{1}{2}$
and spin-$\frac{3}{2}$ polarons. The $\frac{3}{2}$ polaron has
a local spin-1 fluctuation surrounding a spin-$\frac{1}{2}$ core.
The spectral weight for electron removal is exactly zero at in
the region of $k=0$ and $(\pi,\pi)$ because the $\frac{3}{2}$
band crosses below the $\frac{1}{2}$ band.
The spectral weight is also exactly zero at $k=(0,\pi)$ even though
there is no band crossing. We found that this is due to the
orthogonal symmetries of the lowest $k=(0,\pi)$ state and the state
created by removing a $k=(0,\pi)$ electron from the AFM GS.
These findings prompt a more general study about more doping holes; here we
present results for a large cluster with two holes.

The numerical modeling of multi-hole systems at zero temperature is
challenging due to the combinatorially large Hilbert space and the
fermion sign problem. In this work, we further validate a recently proposed
numerical scheme \cite{octapartite} as an efficient and systematic way
of modeling the doped AFM relevant to the lightly doped regime. We
then use it to
calculate the explicit wavefunctions for two  holes introduced
into a half-filled cluster of 32 copper and 64 oxygen sites with
periodic boundary condition. The numerical solution points to a
competition between local attractive potential and kinetic energies.

The article is organized as follows: in Section II we introduce our
model starting from a general three-band model, and then specify
the various assumptions used to simplify it. In Section III we provide
computational details and validate our numerical approach. Section IV
contains our results. Section V further discusses the results and also points out features that are expected
to be robust in larger systems. Section VI contains the conclusions.

\begin{figure}[t]
\includegraphics[width=\columnwidth]{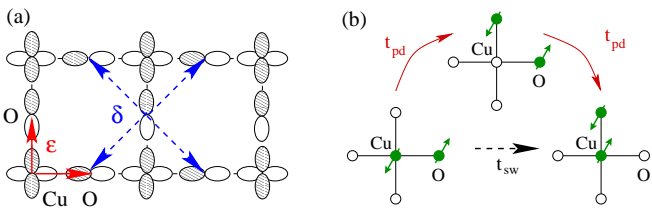}
\caption{\label{fig:cuo2} (color online). (a) Two unit cells of the CuO$_2$
  plane. The orbitals kept in the three-band model of Eq. (\ref{eq:3b}) are
  shown, with white (shaded) for positive (negative) signs. The two
  $\epsilon$ vectors (solid arrow) and
  four $\delta$ vectors (dashed arrow) are also shown. (b) Sketch of
a virtual process of $T_{\rm swap}$.}
\end{figure}

\section{The model}

We start from a three-band $p-d$ model which describes the basic
physics of a hole-doped, charge-transfer gap, insulating
spin-${1\over 2}$ antiferromagnet:\cite{threeband0,threeband1,thomale0,thomale1}
\begin{eqnarray}
H_{3B}&=&T_{pd}+T_{pp} + \Delta_{pd}\sum
n_{l+\epsilon,\sigma}\nonumber\\ &&+U_{pp}\sum
n_{l+\epsilon,\uparrow}n_{l+\epsilon,\downarrow} +U_{dd}\sum
n_{l,\uparrow}n_{l,\downarrow}.\label{eq:3b}
\end{eqnarray}
Here,  $n_{l,\sigma}=d^\dag_{l,\sigma} d_{l,\sigma}$ and
$n_{l+\epsilon,\sigma}=p^\dag_{l+\epsilon,\sigma}
p_{l+\epsilon,\sigma}$ count holes with spin $\sigma$ in the Cu $3d_{x^2-y^2}$ orbital at site $l$, respectively
the O $2p_{x/y}$ orbital located at $l+\epsilon_{x/y}$, and $U_{dd}>U_{pp}>\Delta_{pd}$
describe Hubbard and charge-transfer interactions.
$$
T_{pd}=t_{pd}\sum [(-p^\dag_{l+\epsilon,\sigma}+p^\dag_{l-\epsilon,\sigma}) d_{l,\sigma}
  +h.c.]
$$
and
$$
T_{pp}=t_{pp}\sum s_\delta
p^\dag_{l+\epsilon+\delta,\sigma} p_{l+\epsilon,\sigma}
$$
describe Cu-O, respectively O-O hopping, where the sign
$s_\delta=\delta_x\delta_y/|\delta_x\delta_y|$
of the hopping matrix elements is determined by the relative phases
of the initial and final orbitals. The meaning of the vectors
$\epsilon$ and $\delta$ is explained in Fig. 1(a).
We have taken the direct $T_{dd}$ hoping to be negligible because of
the large spatial separation between Cu$^{2+}$ ions in the lattice. We
have also discarded $T_{pp}'$, the direct next nearest neighbor O--O
hopping, which was found to have negligible effects on the single
polaron states found in
Ref. \onlinecite{spinpolaron} and summarized below in Section II.

At half-filling, in the insulating parent compound, states of mainly
oxygen $2p$ character form a completely filled band and there is one
hole per Cu. If a no-double occupancy restriction is enforced, the
resulting ground-state has AFM order, with a nearest-neighbor Cu-Cu
superexchange interaction mediated by virtual hopping between oxygen
$2p$ and Cu $3d$ orbitals.  An effective model for two doping holes,
located at O sites, interacting with this antiferromagnetic
background, can be derived by direct generalization of the
$U_{dd}\rightarrow\infty$ Rayleigh-Schrodinger method described in the
supplementary material of Ref. \onlinecite{spinpolaron} for the single
hole case. If all hopping integrals are set to zero in the three-band
model (Eq.~\ref{eq:3b}), the two-hole GS are of the type:
\begin{equation}
p_{l+\epsilon,\sigma}^\dag p_{l'+\epsilon',\sigma'}^\dag \prod_{l''}|\sigma_{l''}\rangle_{l''},
\end{equation}
with each ket $|\sigma_{l''}\rangle_{l''}$ specifying the spin of its
Cu. These states have a degeneracy of
$\bigl(\begin{smallmatrix}2N\\2\end{smallmatrix}\bigr)2^{N+2}$. The
  Rayleigh-Schrodinger expansion operates in this highly degenerate
  subspace of the $ N+2$ hole sector of the Fock space to give an
  effective Hamiltonian for these states.  The outcome is slightly
  different than in the single-hole scenario. In particular, the
  projector used in the expansion is
\begin{equation}
P_{2h}=\prod (1-n_{l+\epsilon,\uparrow}
n_{l+\epsilon,\downarrow})\prod(n_{l,\uparrow}+n_{l,\downarrow}-2n_{l,\uparrow}
n_{l,\downarrow}),
\end{equation}
and allows only states with a full lattice of copper spins and no double-occupancies
(due to $U_{pp/dd}$). In other words, the doping holes are forced to
live on O sites. After the expansion, the many resulting terms can be
grouped such that the effective Hamiltonian is written as the sum of two
parts:
\begin{equation}
H_{\rm eff}=P_{2h}H_1P_{2h}+H_2\label{eq:2h_eff}.
\end{equation}
$H_1$ is the collection of terms when the two holes do not directly affect one another.
This single-hole effective Hamiltonian was derived in Ref.~\onlinecite{spinpolaron}:
\begin{equation}
H_1=T_{pp}+T_{\rm swap}+H_{J_{pd}}+H_{J_{dd}},\label{eq:h1}
\end{equation}
where the bare oxygen-oxygen hopping of a hole, $T_{pp}$, is supplemented by:
\begin{eqnarray}
T_{swap}&=&-t_{sw} \sum s_{\eta}p^\dag_{l+\epsilon+\eta,\sigma}p_{l+\epsilon,\sigma'}
|\sigma'_{l_{\epsilon,\eta}}\rangle\langle\sigma_{l_{\epsilon,\eta}}|\label{eq:tex}\\
H_{J_{pd}}&=&J_{pd} \sum\overline{S}_{l}\cdot\overline{S}_{l\pm\epsilon}\label{eq:jpd}\\
H_{J_{dd}}&=&J_{dd}\sum\overline{S}_{l\pm2\epsilon}\cdot\overline{S}_l\Pi_\sigma
(1-n_{l\pm\epsilon,\sigma})\label{eq:jdd}
\end{eqnarray}
The physics described by these terms has been discussed in some detail
in the supplementary
materials of Ref. \onlinecite{spinpolaron}.
Briefly, $H_{J_{pd}}$ is the exchange between the spin of a doping
hole located at an O site, with that of its neighboring Cu spins. This
term favors the formation of a three-spin polaron
(3SP),\cite{emery1,exp3sp} either $|\Uparrow\rangle$ or
$|\Downarrow\rangle$ -- the corresponding eigenfunctions for the
central Cu-O-Cu spins are listed in Table ~\ref{tab:2h_3sp}. These
describe the ferromagnetic core of the hole-induced disturbance in the
otherwise AFM background.  $H_{J_{dd}}$ is the usual superexchange
between neighboring Cu spins, which, however, is blocked if a hole is
located on the ligand O.  Such blocking decreases the penalty for
having the 3SP ferromagnetic core in the AFM background. Finally,
$T_{swap}$ describes the processes sketched in Fig. \ref{fig:cuo2}(b),
where the hole from a Cu neighbor to the doping O hole first hops to
another of its three hole-free O neighbors, followed by the original
hole falling into the Cu orbital.  The opposite phase between $T_{pp}$
and $T_{swap}$ favors the coherent propagation of the three-spin
polarons. Even in the single-hole scenario, the physics encompassed by
these terms leads to qualitative differences when compared to the
t-t'-J model.\cite{spinpolaron}

\begin{table}
\centering
\begin{tabular}{l | c | r}
Wavefunction & Total Spin & $\frac{\langle H_{J_{pd}}\rangle}{J_{pd}}$\\
\hline
$|\Uparrow\rangle=\sqrt{\frac{1}{3}}
p^\dag_{\uparrow}\frac{|\uparrow\downarrow\rangle+|\downarrow\uparrow\rangle}{\sqrt{2}}
-p^\dag_{\downarrow}\sqrt{\frac{2}{3}}|\uparrow\uparrow\rangle $ & $\frac{1}{2}$ & $-1$\\
$|\Downarrow\rangle=\sqrt{\frac{1}{3}}
p^\dag_{\downarrow}\frac{|\uparrow\downarrow\rangle+|\downarrow\uparrow\rangle}{\sqrt{2}}
-p^\dag_{\uparrow}\sqrt{\frac{2}{3}}|\downarrow\downarrow\rangle $ & $\frac{1}{2}$ & $-1$\\
$|0+\rangle=\sqrt{\frac{1}{3}}
p^\dag_{\uparrow}\frac{|\uparrow\downarrow\rangle-|\downarrow\uparrow\rangle}{\sqrt{2}}
$ & $\frac{1}{2}$ & $0$\\
$|0-\rangle=\sqrt{\frac{1}{3}}
p^\dag_{\downarrow}\frac{|\uparrow\downarrow\rangle-|\downarrow\uparrow\rangle}{\sqrt{2}}
$ & $\frac{1}{2}$ & $0$\\
$|\frac{3}{2},\frac{3}{2}\rangle=p^\dag_{\uparrow}|\uparrow\uparrow\rangle
$ & $\frac{3}{2}$ & $\frac{1}{2}$\\
$|\frac{3}{2},\frac{1}{2}\rangle=\sqrt{\frac{2}{3}}
p^\dag_{\uparrow}\frac{|\uparrow\downarrow\rangle+p^\dag_{\downarrow}|\downarrow\uparrow\rangle}{\sqrt{2}}
+
\sqrt{\frac{1}{3}}p^\dag_{\downarrow}|\uparrow\uparrow\rangle$ & $\frac{3}{2}$ & $\frac{1}{2}$\\
$|\frac{3}{2},-\frac{1}{2}\rangle=\sqrt{\frac{2}{3}}
p^\dag_{\downarrow}\frac{|\uparrow\downarrow\rangle+|\downarrow\uparrow\rangle}{\sqrt{2}}
+
\sqrt{\frac{1}{3}}p^\dag_{\uparrow}|\downarrow\downarrow\rangle $ & $\frac{3}{2}$ & $\frac{1}{2}$\\
$|\frac{3}{2},-\frac{3}{2}\rangle=p^\dag_{\downarrow}|\downarrow\downarrow\rangle
$ & $\frac{3}{2}$ & $\frac{1}{2}$
\end{tabular}
\caption{\label{tab:2h_3sp} Single-hole eigenstates of
  $H_{J_{pd}}$. $p^\dag_\sigma$ creates an oxygen hole and the arrows
  in the ket indicate the spins of the two copper sites neighboring
  the oxygen hole.}
\end{table}

Now we identify the important terms contributing to $H_2$, which
describes the evolution of the two doping holes when close to each
other.  We write:
\begin{equation}
H_2=H_2^{(2)}+H_2^{(3)}+H_2^{(4)}\cdots
\end{equation}
keeping up to fourth-order terms in the Rayleigh-Schrodinger
expansion. In the expansion, the appearance of $(1-P_{2h})$ dictates
that all terms  must have an even power of $t_{pd}$ because
the final states must have one copper spin per unit cell. Because
$H_2$ accounts for only 2-hole corrections to $P_{2h}H_1P_{2h}$, one
can deduce that all terms in the second-order $H_2^{(2)}$ share a
factor of $t_{pp}^2$. All terms in the third-order $H_2^{(3)}$ share a
$t_{pd}^2t_{pp}$ factor because the second step of any three-step
$t_{pp}^3$ process would yield virtual excited states projected out by
$(1-P_{2h})$. The fourth-order $H_2^{(4)}$ collects terms proportional
to $t_{pp}^4$, $t_{pp}^2t_{pd}^2$ or $t_{pd}^4$.

The second order corrections are $\frac{t_{pp}^2}{U_{pp}}$ processes that link
initial and final states in which oxygen holes are $\delta$ apart
(Fig. 1a). Because we are considering only the $\sigma$-bonding  oxygen orbitals, the virtual
excitation is a state with double occupancy of an oxygen orbital; that is, the matrix
element is non-zero only for singlet correlations:
\begin{widetext}
\begin{equation}
H_2^{(2)}=\frac{2t_{pp}^2}{U_{pp}}\sum
(-1)^{\delta(\delta_1\cdot\delta_2)}\left(\overline{S}_{\alpha'\alpha}\cdot\overline{S}_{\beta'\beta}-\frac{1}{4}\right)
p^\dag_{l+\epsilon+\delta_1,\alpha'}
p_{l+\epsilon+\delta_1,\alpha}p^\dag_{l+\epsilon+\delta_1+\delta_2,\beta'}
p_{l+\epsilon,\beta},
\end{equation}
\end{widetext}
with
$$\overline{S}_{\alpha'\alpha}\cdot\overline{S}_{\beta'\beta}=\frac{1}{4}\sum_{i=x,y,z}\sigma^i_{\alpha'\alpha}\sigma^i_{\beta'\beta},$$
where $\sigma^i_{\alpha'\alpha}$ is an element of the pauli matrix in the i$^{th}$ direction.
$\delta_{1,2}$ sums over all generic $\delta$ values (Fig.~\ref{fig:cuo2}a). $\delta(\delta_1\cdot\delta_2)=0[1]$ if $\delta_1$ and $\delta_2$ are parallel [perpendicular].
 The
matrix element is non-zero only for singlet-like configuration and when the two holes are
$|\delta|=\frac{a}{\sqrt{2}}$ apart. There are 8 non-zero matrix
elements in this situation. Two of these correspond to
$\delta_1+\delta_2=0$ so the static AFM exchange is
$2\times\frac{2t_{pp}^2}{U_{pp}}$. There are also 6 other ways for one
hole to ``skip" over the other with such a Heisenberg factor.

There is an abundance of terms in third- and fourth-order,
which are a sub-set of terms studied in the literature of 2-band
models.\cite{zaanen2band0,zaanen2band1,zaanen2band2,zaanen2band3,zaanen2band4}
To provide a simple physical picture, we consider only terms that are
greater or equal to $J_{dd}$, which is roughly $\frac{8\cdot16\cdot
  t_{pp}^4}{3U_{pp}^3}\sim \frac{16}{81}t_{pp}$ for $t_{pd}\sim
2t_{pp}$ and $\Delta_{pd}\sim U_{pp}\sim 6t_{pp}$. The basis of this
approximation is the observation that the dominant short-range physics
should be already captured by the numerous low-order terms with an
energy scale of $t_{pp}$ and $T_{swap},J_{pd},J_{pp} \sim 0.66t_{pp}$,
which are already greater than the long-range physics at the scale of
$J_{dd}\sim 0.2 t_{pp}$. Adding the relevant short-range corrections
with magnitude greater than $0.2 t_{pp}$ should then be more than
adequate. We also discards terms that can be factored into
$T_{pp},T_{swap},J_{pd},J_{pp}$ and the identity processes. These
terms effectively scale  processes by some overall factors which
roughly cancel out when all parameters are divided by $J_{dd}$
in order to use
dimensionless parameters, since $J_{dd}$
also undergoes similar renormalizations in higher order.

As discussed above, all third-order corrections have a prefactor of $t_{pp}t_{pd}^2$.
The perturbation goes through two virtual states so the largest possible magnitude is
$\frac{t_{pp}t_{pd}^2}{\Delta_{pd}U_{pp}}\sim
\frac{t_{pp}t_{pd}^2}{\Delta_{pd}\Delta_{pd}}\sim \frac{1}{9}
t_{pp}$.
The splitting due to such a pair of Hermitian matrix elements is $\sim
\frac{2}{9}t_{pp}$, of the  same order as the $J_{dd}$ splitting. These processes
involve virtual excitations with no double occupancy. They are all multiples of
$t_{pp},T_{swap},J_{pd}$ and the identity processes and are
discarded due to the rescaling argument discussed above. Because
$U_{pp}\sim \Delta_{pd}$,
other processes would have denominators that are at least a factor of
two greater; that is,
constructive quantum interference is required for any terms to be non-negligible
compared to the superexchange. Constructive interference among $t_{pp}t_{pd}^2$ processes
requires the same orbital occupancy in the initial and final state, and this can happen
only if the two oxygen holes are $|\delta|$ apart. The transition of interest
is then $p^\dag_{l\pm e_x,\alpha}p^\dag_{l\pm
  e_y,\beta}d^\dag_{l,\gamma}\rightarrow p^\dag_{l\pm
  e_x,\alpha'}p^\dag_{l\pm e_y,\beta'}d^\dag_{l,\gamma'}$. The
effective matrix elements have the form
$\langle\alpha',\beta',\gamma'|H^{(3)}_2|\alpha,\beta,\gamma\rangle$, a local three-spin
ring involving the copper spin sandwiched by the oxygen holes. The
correction term can be
expressed as a summation over each copper spin with two additional vector $\Delta_{x/y}$
summed over $\pm\epsilon_{x/y}$. Noting that the 3-step hopping would give an overall
phase of $-1$ (Fig.~\ref{fig:cuo2}a) and that the virtual states
have double-occupancy, it is not
surprising that all operators contain a shifted Heisenberg factor:

\begin{widetext}
\begin{eqnarray}
H_2^{(3)}=\frac{2t_{dp}^2t_{pp}}{U_{pp}(\Delta_{pd}+U_{pp})}\sum&&
\left(\overline{S}_{\beta'\beta}\cdot\overline{S}_{\alpha'\alpha}-\frac{1}{4}\right)p^\dag_{l+\Delta_x,\beta'} p_{l+\Delta_x,\alpha}p^\dag_{l+\Delta_y,\gamma} p_{l+\Delta_y,\beta}|l,\alpha'\rangle\langle l,\gamma|\nonumber\\
&+&
\left(\overline{S}_{\alpha'\alpha}\cdot\overline{S}_{\beta'\beta}-\frac{1}{4}\right)p^\dag_{l+\Delta_x,\gamma} p_{l+\Delta_x,\alpha}p^\dag_{l+\Delta_y,\alpha'} p_{l+\Delta_y,\beta}|l,\beta'\rangle\langle l,\gamma|\nonumber\\
&+&\left(\overline{S}_{\gamma'\gamma}\cdot\overline{S}_{\alpha'\alpha}-\frac{1}{4}\right)p^\dag_{l+\Delta_x,\gamma'} p_{l+\Delta_x,\alpha}p^\dag_{l+\Delta_y,\alpha'} p_{l+\Delta_y,\beta}|l,\beta\rangle\langle l,\gamma|\nonumber\\
&+&\left(\overline{S}_{\beta'\beta}\cdot\overline{S}_{\gamma'\gamma}-\frac{1}{4}\right)p^\dag_{l+\Delta_x,\beta'} p_{l+\Delta_x,\alpha}p^\dag_{l+\Delta_y,\gamma'} p_{l+\Delta_y,\beta}|l,\alpha\rangle\langle l,\gamma|\nonumber\\
+\frac{2t_{dp}^2t_{pp}}{(\Delta_{pd}+U_{pp})^2}\sum&&\delta_{\alpha\beta}\left(\overline{S}_{\beta'\beta}\cdot\overline{S}_{\gamma'\gamma}-\frac{1}{4}\right)p^\dag_{l+\Delta_x,\alpha} p_{l+\Delta_x,\alpha}p^\dag_{l+\Delta_y,\beta'} p_{l+\Delta_y,\beta}|l,\gamma'\rangle\langle l,\gamma|\nonumber\\
&+&\delta_{\gamma\beta}\left(\overline{S}_{\alpha'\alpha}\cdot\overline{S}_{\beta'\beta}-\frac{1}{4}\right)p^\dag_{l+\Delta_x,\gamma} p_{l+\Delta_x,\alpha}p^\dag_{l+\Delta_y,\alpha'} p_{l+\Delta_y,\beta}|l,\beta'\rangle\langle l,\gamma|\nonumber\\
&+&\delta_{\gamma\alpha}\left(\overline{S}_{\beta'\beta}\cdot\overline{S}_{\alpha'\alpha}-\frac{1}{4}\right)p^\dag_{l+\Delta_x,\beta'} p_{l+\Delta_x,\alpha}p^\dag_{l+\Delta_y,\gamma} p_{l+\Delta_y,\beta}|l,\alpha'\rangle\langle l,\gamma|\nonumber\\
&+&\delta_{\beta\alpha}\left(\overline{S}_{\alpha'\alpha}\cdot\overline{S}_{\gamma'\gamma}-\frac{1}{4}\right)p^\dag_{l+\Delta_x,\alpha'} p_{l+\Delta_x,\alpha}p^\dag_{l+\Delta_y,\beta} p_{l+\Delta_y,\beta}|l,\gamma'\rangle\langle l,\gamma|
\end{eqnarray}
\end{widetext}

These terms are finite if there is a 2-spin singlet  amongst the 3 spins.
The first four terms give an eigenvalue of
$+4\frac{t_{dp}^2t_{pp}}{U_{pp}(\Delta_{pd}+U_{pp})}\sim 0.2t_{pp}$
for oxygen-oxygen singlet pair and $0$ for triplet pair. The last four terms give
an eigenvalue of $\frac{t_{dp}^2t_{pp}}{(\Delta_{pd}+U_{pp})^2}\sim 0.02t_{pp}$
for singlet pairs and $0,-3\frac{t_{dp}^2t_{pp}}{(\Delta_{pd}+U_{pp})^2}\sim -0.06t_{pp}$
for triplets. Therefore we can approximate this term by simply raising the energy of
oxygen-oxygen singlet pairs accordingly.

In the fourth-order, all $t_{pd}^4$ processes can be factored into two
$T_{swap}$ or $J_{pd}$ processes. The $t_{pd}^2t_{pp}^2$ processes are smaller than
$J_{dd}$ by a factor of $\sim4\times 4$ and no constructive interference is possible.
$2t_{pp}^4$ processes are even smaller. Therefore, we set $H_2^{(4)}\approx 0$.
In summary, the 2-hole correction is in essence:

\begin{widetext}
\begin{eqnarray}
H_2\approx&+&J_{pp}^{(2)}\sum (-1)^{\delta(\delta_1\cdot\delta_2)}\left(\overline{S}_{\alpha'\alpha}\cdot\overline{S}_{\beta'\beta}-\frac{1}{4}\right)p^\dag_{l+\epsilon+\delta_1,\alpha'} p_{l+\epsilon+\delta_1,\alpha}p^\dag_{l+\epsilon+\delta_1+\delta_2,\beta'} p_{l+\epsilon,\beta}\nonumber\\
&-&J_{pp}^{(3)}\sum\left(\overline{S}_{\alpha'\alpha}\cdot\overline{S}_{\beta'\beta}-\frac{1}{4}\right)p^\dag_{l+\epsilon+\delta,\alpha'} p_{l+\epsilon+\delta,\alpha}p^\dag_{l+\epsilon,\beta'} p_{l+\epsilon,\beta}\label{eq:2h_h2}
\end{eqnarray}
\end{widetext}
with $J_{pp}^{(2)}=\frac{2t_{pp}^2}{U_{pp}}$ and $J_{pp}^{(3)}=\frac{4t_{dp}^2t_{pp}}{U_{pp}(\Delta_{pd}+U_{pp})}$ for hole-hole interaction due to second and third order corrections.
Using $t_{pd}=1.3eV$,
$t_{pp}=0.65eV$, $\Delta_{pd}=3.6eV$, and
$U_{pp}=4eV$,\cite{tjrev0} we scale the parameters in units of
$J_{dd}$ to find their dimensionless values to be $t_{pp}=4.13$,
$t_{sw}=2.98$, $J_{pd}=2.83$, and $J_{pp}^{(2)}=1.3420$, and $J_{pp}^{(3)}=0.9182$.

\section{The computation}

Although the thermodynamic limit is important in condensed
matter theory, models detailing the interactions between fermionic
carriers and a quantum antiferromagnet present formidable challenges
to analytical solutions. Finite cluster numerical calculations are
therefore a valuable tool in extracting information about the model of
interest. Numerical studies of phase transitions would require
$N\rightarrow\infty$ extrapolation; instead, this work focuses on the
correlations between two holes injected in a half-filled $N=32$
CuO$_2$ cluster with superior topological properties compared to
smaller clusters.\cite{betts} The computational breakthrough described
in this section allows us to explicitly obtain the low-energy
wavefunctions. We can then calculate any correlators of interest, in
order to understand the nature of these eigenstates. We can also
perform a direct comparison with the results of the t-t'-J model with
two holes on the same $N=32$ cluster.\cite{2hed}

\subsection{An insight about antiferromagnets}
Even after exploiting translational and spin-projection symmetries,
two holes injected into a half-filled $N=32$ CuO$_2$ cluster have
a Hilbert space with $0.154\times10^{12}$ states. A system of this
size challenges the capability of all unbiased methods at zero
temperature. One way forward is to identify how to drastically reduce this
dimension while keeping the most important states in the basis. The
two doping holes contribute at most a factor of
$4\bigl(\begin{smallmatrix}2N\\2\end{smallmatrix}\bigr)$, so
it is the AFM background that is primarily responsible for this large
number.

We have previously proposed an efficient numerical approach for
the modeling of  antiferromagnets,\cite{octapartite} upon which we will build here.
For completeness, we first briefly review the undoped
scenario\cite{octapartite} -- here
the system is described by AFM Heisenberg superexchange between neighbor Cu
spins, see Eq. (\ref{eq:jdd}). If the Cu
spins  on the square lattice
 are divided into two sublattices, A and B, such that
spins from each sublattice only couple to those of the
other sublattice, one of the many measures of the  staggered magnetization
can be expressed in terms of the total spin $S=S_A+S_B$:
\begin{equation}
\hat{m^2}=\frac{1}{N^2}\left(\sum_r (-1)^{|r|} \hat{S}_{r}\right)^2 =
\frac{1}{N^2}(2\hat{S}_{A}^2+2\hat{S}_{B}^2-\hat{S}^2).
\label{eq:sm_dev}
\end{equation}
The undoped GS is known to be a singlet, $S=0$. For $\hat{S}_{A}$ and
$\hat{S}_{B}$ to add to zero, their quantum numbers must be equal,
$S_A=S_B$. Moreover, there are
accurate estimates of $m$, which range from $\sim0.3$ as
$N\rightarrow \infty$ to $\sim 0.45$ for $N=32$.\cite{Mans}
In order to get such large values for $m$,  the sublattice spins ${S_{A/B}}$
have to be within $\frac{N}{16}$ of
their maximum values. In other words, in each sublattice $\frac{N}{8}$ spins
add to a total spin of zero while the rest add to the maximum $3\frac{N}{16}$.
Based on this insight, we showed that the GS can be captured
systematically by considering a subspace
specified by a completeness parameter
\begin{equation}
C_S\in [0,1].
\end{equation}
The subspace equals
the full Hilbert space when $C_S=1$. For smaller values of $C_S$, only the states
with
\begin{equation}
S_{A/B}\geq \frac{N}{4}\left(1-C_S\right)\label{eq:truncation}
\end{equation}
are included. $C_S=0$ is, therefore, a singlet containing the classical N\'eel state.
A linear decrease of $C_S$  yields a combinatorial decrease in the number of states.
In Ref. \onlinecite{octapartite} we postulated apriorily that the
subspace with $C_S=\frac{1}{4}$
captures the essence of the
wavefunction, and then showed that  the convergence
is exponential as $C_S$ is tuned from zero to unity and  that
$C_S=\frac{1}{4}$ is indeed
the ``sweet spot".

For a large sample, two additional holes
cannot drastically change the entire AFM background;
therefore, the stable, systematic convergence of the undoped
AFM is expected to hold for the two-hole scenario.

\begin{figure}[t]
\centering
\includegraphics[width=0.8\columnwidth]{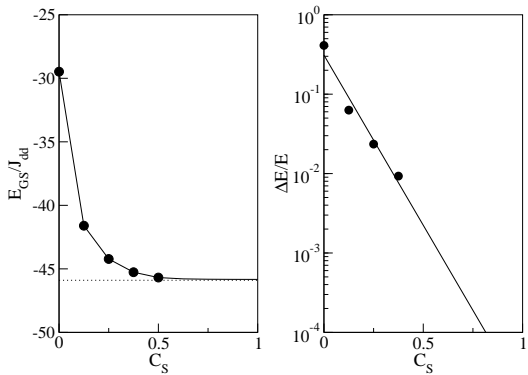}
\caption{\label{fig:bpFig1} Convergence of the one-hole ground state
  energy. Left: GS energy calculated for increasing $C_S$. The value
approaches rapidly the exact value marked by the dotted
line. Right: fractional change in the GS energy for the next increment of
$C_s$. Solid lines are extrapolation and linear fits. }
\end{figure}

\begin{table}[]
\centering
\begin{tabular}{c | r r r r r r | r r r}
  $_{S_A} \backslash ^{S_B}$ & 0 & 1 & 2 & 3 & 4 & 5 & 6 & 7 & 8 \\
  \hline
  0 & 0.00 & 0.00 & 0.00 & 0.00 & 0.00 & 0.00 & 0.00 & 0.00 & 0.00 \\
  1 & 0.00 & 0.02 & 0.03 & 0.00 & 0.00 & 0.00 & 0.00 & 0.00 & 0.00 \\
  2 & 0.00 & 0.03 & 0.14 & 0.16 & 0.00 & 0.00 & 0.00 & 0.00 & 0.00 \\
  3 & 0.00 & 0.00 & 0.16 & 0.69 & 0.68 & 0.00 & 0.00 & 0.00 & 0.00 \\
  4 & 0.00 & 0.00 & 0.00 & 0.68 & 2.52 & 2.21 & 0.00 & 0.00 & 0.00 \\
  5 & 0.00 & 0.00 & 0.00 & 0.00 & 2.21 & 7.06 & 5.28 & 0.00 & 0.00 \\
    \hline
  6 & 0.00 & 0.00 & 0.00 & 0.00 & 0.00 & 5.28 & 14.24 & 8.44 & 0.00 \\
  7 & 0.00 & 0.00 & 0.00 & 0.00 & 0.00 & 0.00 & 8.44 & 18.06 & 6.85 \\
  8 & 0.00 & 0.00 & 0.00 & 0.00 & 0.00 & 0.00 & 0.00 & 6.85 & 9.97 \\
\end{tabular}
\caption{\label{tab:sp32_05_05_weight} (N=32) The S=$\frac{1}{2}$
  $k=(\frac{\pi}{2},\frac{\pi}{2})$ single-hole ground state's probability
  in subspaces of particular $S_A \bigotimes S_B$ values. Numbers are
  percentages adding up to 100\%. Two lines are drawn to highlight the
  $C_S=\frac{1}{4}$ truncation which
discards states with $S_A,S_B<6$ and yields an energy within 3.6\% of the exact value.}
\end{table}

We first test the method against the exactly solvable scenario of one
hole on a cluster with 32 Cu and 64 O.\cite{spinpolaron} Using the
same truncation criterion, we calculate the one-hole ground state for
increasing $C_S$. The convergence is illustrated in
Fig.~\ref{fig:bpFig1}.  The energy computed for $C_S=\frac{1}{4}$ and
$\frac{1}{2}$ is within 3.6\% and 0.5\%, respectively, of the exact
 value. Table \ref{tab:sp32_05_05_weight} shows the exact
wavefunction's probability in various subspaces $S_A \bigotimes S_B$ of the
sublattice total spins $S_{A/B}$. While the $C_S=\frac{1}{4}$
truncation captures $\sim$ 95\% of the probability of the undoped
wavefunction,\cite{octapartite} Table \ref{tab:sp32_05_05_weight}
shows that $C_S=\frac{1}{4}$ captures $\sim$ 73\% of the one-hole
groundstate.  The next two increments of $C_S$ contains 17\% and 7\%,
respectively, of the remaining weight.  The addition of a hole couples
the spin background to adjacent values of $S_A \bigotimes S_B$, and
states added by increasing $C_S$ have decreasing importance in the
wavefunction; therefore, it is reasonable to expect an increment of
$\delta C_S \sim \frac{1}{N}$ from $C_S=\frac{1}{4}$ to suffice. At
least in the very dilute limit, the truncation scheme provides a good
starting point to systematically capture the low energy state.

These observations provide reasonable merits for the application of
this scheme to the two-hole scenario, whose ground state should be
captured in the subspace of $C_S\sim \frac{1}{4}+O(\delta C_S)$ for
large $N$. To provide a conservative error analysis,
however, we would aim for a capability of up to $C_S\sim \frac{1}{2}$,
which in turn limits the cluster size to $N=32$.  We showed in our
previous work\cite{spinpolaron} that a single oxygen hole  induces a
local disturbance in the AFM background and that, for the parameter
range of interest, the disturbance affects $\sim 6-12$ spins around the
hole. Therefore, a cluster with 32 unit cells should be large enough to
accommodate two holes without artificially forcing them too close
together. As discussed below, our results are different from those yielded by
similar attempts restricted to $N=16$ unit
cells.\cite{zaanen2band0,zaanen2band1,zaanen2band2,zaanen2band3,zaanen2band4}

\subsection{Implementation}
The Hilbert space of the two-hole  problem (Eq.~\ref{eq:2h_eff}
together with Eqs.~\ref{eq:h1} and \ref{eq:2h_h2}) contains
$\bigl(\begin{smallmatrix}2N\\2\end{smallmatrix}\bigr)2^{N+2}$ states of
the form $p_{l+\epsilon,\sigma}^\dag p_{l'+\epsilon',\sigma'}^\dag
\prod|\sigma_{l''}\rangle$, with $l+\epsilon\neq l'+\epsilon'$.
To perform computations for large $N$, this basis must be reorganized to take
advantage of translational symmetry, total-spin symmetry, total-spin-projection
symmetry, and, most importantly, the truncation scheme discussed in
the previous section.
The implementation  is detailed in
Appendix A.

We emphasize that the truncation scheme is applied only to
the AFM background, i.e. to the parts of the wavefunction describing
the spins at Cu sites. There is no
restriction for the two doping holes residing on the O sites, apart from
no-double occupancy.

\section{Results}

We apply the approach described above to solve the two doping holes
problem for the $N=32$ cluster. We find that the low-energy states
have a total spin of $S_T=0$ or $S_T=1$ (the total spin includes both
the contribution of the AFM background and of the doping holes).
Sectors with higher $S_T$ have higher energies, in accordance with the
trend of finite size AFM computations.\cite{Mans} As discussed in
Ref. \onlinecite{spinpolaron}, the one-hole GS are degenerate
at $k=(\pm\frac{\pi}{2},\pm\frac{\pi}{2})$, so the important total
momenta for two holes are $K=(0,0),(\pi,0),(\pi,\pi)$. The convergence of the lowest
states at these high-symmetry points is shown in
Fig.~\ref{fig:bpFig3}. The trend of exponential convergence is similar
to that found for the undoped\cite{octapartite} and single hole cases
(Fig.~\ref{fig:bpFig1}), signaling that the dominant part of the
Hilbert space has been captured. We first present the energetics then
illustrate the numerical eigenvectors in details.

\subsection{Energetics}

\begin{figure}[t]
\centering
\includegraphics[width=0.8\columnwidth]{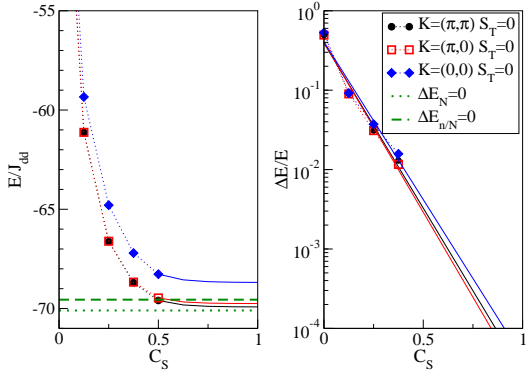}
\caption{\label{fig:bpFig3} Convergence of two-hole low energy states
  at several high-symmetry points. The horizontal dashed and dotted lines are the
  $\Delta E_{n/N}=0$ and $\Delta E_{N}=0$ levels, respectively, i.e.
  zero binding energy levels calculated at fixed doping concentration and
  fixed lattice size, respectively.
  The two values bound the binding energy value and converge as $N\rightarrow\infty$.}
\end{figure}

\begin{figure}[t]
\centering
\includegraphics[width=0.8\columnwidth]{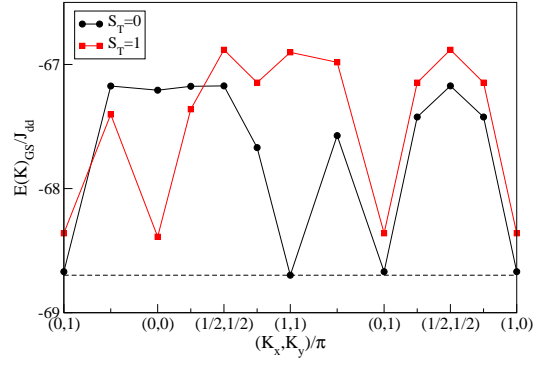}
\caption{\label{fig:bpFig2} Energy of the lowest two-hole states vs total momentum
  along high-symmetry cuts, for $S_T=0,1$  and
  $C_S=\frac{3}{8}$.}
\end{figure}

Figure~\ref{fig:bpFig2} shows the dispersion of the lowest two-hole
states versus the  total momentum $K$, for  $S_T=0,1$. In the
$S_T=0$ sector, the $K=(\pi,\pi)$ global groundstate is doubly
degenerate. The $K=(\pi,0)$ and $K=(0,\pi)$ states are $\sim
0.17J_{dd}$ higher in energy, while the $K=(0,0)$ state is almost $2J$
higher in energy. The shape of the energy dispersion
mirrors that of some previous studies treating $N\ge 32$ systems with
two holes.\cite{2hed,berciu} Although finite-momentum superconducting pair has been proposed by Fulde, Ferrell, Larkin, and Ovchinnikov\cite{fflo0,fflo1}, this scenario is not superconducting.
For $N=16$, the GS is instead found to be located at
$K=(\pi,0)$.\cite{zaanen2band4} This signals
that $N=16$ is definitely too small to capture the two-hole physics,
which is not surprising since
a single polaron was found to disturb $\sim6-12$ spins in its
vicinity.\cite{spinpolaron}

As for the $S_T=1$ sector, these states cross below the $S_T=0$ states
in the region around $K=0$. The lowest $S_T=1, K=0$ level is doubly
degenerate as well. Its energy is $\sim
0.3J_{dd}$, equal to the $\sim\frac{1}{N}$ free magnon gap of
the undoped $N=32$ cluster.\cite{Mans} The following section confirms that
these  $S_T=1, K=0$ low energy states have  the two holes in their doubly
degenerate global GS with $S_T=0, K=(\pi,\pi)$, plus
a $Q=(\pi,\pi), S=1$
``magnon''-like excitation in the AFM
background.  It is thus
reasonable to expect the $S_T=1, K=0$ states to become degenerate with
the $K=(\pi,\pi), S_T=0$ states as $N\rightarrow\infty$ and the free
magnon gap closes. Similarly, the low energy $S_T=1$ excitations at
$K=(\pi,0)$ and $K=(0,\pi)$ are the $S_T=0$ two-hole state at
$K=(0,\pi)$, respectively $K=(\pi,0)$, plus a $Q=(\pi,\pi), S=1$
``magnon''. They are also expected to become degenerate as
$N\rightarrow \infty$.

For the one-hole case, we were able to establish the robustness of
spin-1 excitations because we found that $E_{3/2}-E_{1/2}$ is much
lower than any finite-size spin excitation,
$E_{magnon}-E_{1/2}$.\cite{spinpolaron} In the two-hole case, however, all
spin excitations are higher than $E_{magnon}-E_{0}$, and no conclusions can be drawn.

We note that the two-hole bandwidth is $\sim 2J_{dd}$, which is roughly equal to the single-polaron bandwidth.\cite{spinpolaron} This suggests that the GS has very weak, if any, binding. (The $\sim 2J_{dd}$ scale has further implication when the wavefunction is considered in Sec.~IV.) The binding energy is the difference between the two-hole energy and twice the one-hole
energy, shifted by the energy of the undoped system:
\begin{equation}
\Delta E= (E_{2h}-E_{0h})-2(E_{1h}-E_{0h}).
\end{equation}
This value must be extracted with care for a finite cluster. Although
the $N=32$ cluster can accommodate two spin polarons, and each disturbs
$\sim 6-12$ Cu spins in its vicinity,\cite{spinpolaron} as we show in
the next section  we find that
the low energy states have a low probability to
be close enough to share a common Cu. Unlike the
$N\rightarrow\infty$ limit, the local range of this blockade is not
negligible in a $N=32$ cluster so the kinetic energy is
over-estimated. The one-hole scenario does not have this artifact,
leading to an overestimation if the value is evaluated from energies
calculated at a constant $N$,
\begin{equation}
\Delta E_{N} = (E^{(32)}_{2h}-E^{(32)}_{0h})-2(E^{(32)}_{1h}-E^{(32)}_{0h}).
\end{equation}

Given that $ N$-dependent scaling is impossible because
$N=16$ clusters have a different ground state (e.g. at momentum
$(0,\pi)$ instead of $(\pi,\pi)$),\cite{zaanen2band4} and that a larger
cluster cannot currently be solved,
we estimate a reasonable lower bound for the binding energy by considering
the one- and two-hole energies at fixed concentration
$\frac{n}{N}=\frac{2}{32}=\frac{1}{16}$,
\begin{equation}
\Delta E_{\frac{n}{N}} = (E^{(32)}_{2h}-E^{(32)}_{0h})-2(E^{(16)}_{1h}-E^{(16)}_{0h}).
\end{equation}
These  definitions converge in the large $N$ limit,
\begin{equation}
\Delta E = \lim_{N\rightarrow\infty} \Delta E_{N}=\lim_{N\rightarrow\infty} \Delta E_{\frac{n}{N}},
\end{equation}
and thus give the upper- and lower-bound for the $N=32$ cluster.

As shown in Fig.~\ref{fig:bpFig3}, the GS energy was extrapolated to be$E^{(32)}_{2h}=-69.917J_{dd}$. The $\Delta E_{N}=0$ and $\Delta E_{\frac{n}{N}}=0$ levels in the same figure indicate a binding energy of $\left(-0.091\pm 0.272\right)J_{dd}$. This suggests weak, if any, binding, although solutions for larger clusters are required to reduce the error bars sufficiently to be able to draw a rigorous conclusion.

\subsection{Wavefunction analysis}

\subsubsection{Symmetry}

We define $P_{xy/\underline{x}y}$ for reflections about the diagonals.
The two degenerate $S=0,$ $K=(\pi,\pi)$ GSs were found to have
$P_{xy}=-P_{\underline{x}y}=\pm 1$. The two degenerate $S=1,$
$K=(0,0)$ states related to them by a gapless
$Q=(\pi,\pi)$ magnon were also found to have $P_{xy}=-P_{\underline{x}y}=\pm 1$.
This is $p$-wave parity, but it is not a concern because we are outside of the
superconducting region. For example, the product of two
$P_{xy}=-P_{\underline{x}y}=\pm 1$ wavefunctions would yield
a $d_{x^2-y^2}$ symmetry. Defining $P_{x/y}$ for reflections about
the lattice parameter directions, the $K=(0,\pi)$ state was found to have
$p_y$ symmetry and $K=(\pi,0)$ state has $p_x$ symmetry as in the t-t'-J model.

Unlike in the $N=32$ t-t'-J model whose lowest two-hole
$S=0, K=0$ state has s-symmetry\cite{2hed}, the lowest $S=0, K=0$ state has
$P_{xy}=P_{\underline{x}y}=-1$, which is $d_{x^2-y^2}$
symmetry. Note that the lowering of s-symmetric state in the t-t'-J model is due
to the t' hopping of the Zhang-Rice singlet, and the t' hopping is included in
this work due to the explicit consideration of oxygen-oxygen hopping.

\subsubsection{Charge correlations}

The charge correlation as a function of the separation $R$ between the
two doping holes can be
characterized by the expectation value of:
\begin{eqnarray}
\hat{c}(\eta) &=&\sum_{l,\epsilon,\sigma,\sigma'} n_{l+\epsilon,\sigma}n_{l+\epsilon+\eta,\sigma'}\\
\hat{C}(R) &=& \sum_{|\eta|=R} \hat{c}(\eta)
\end{eqnarray}
with $\sum_R \hat{C}(R)=1$.
On a finite cluster, the number of hole-hole configuration separated by a distance
$R>\frac{L}{2}$ is limited by the periodic boundary condition to be
smaller than in the
$L^2=N\rightarrow\infty$ limit. Therefore, such correlations should be
compared to the probability $P(R)$ of two randomly distributed oxygen
holes to be separated by $R$ in the same finite cluster. Then, the
correlations can be gauged by
\begin{equation}
\Delta \hat{C}(R)=\hat{C}(R)-P(R),
\end{equation}
meaning that the correlation is the same as that of a random
distribution if $\Delta C(R)=\langle\Delta \hat{C}(R)\rangle=0$.

\begin{figure}[t]
\centering
\includegraphics[width=0.8\columnwidth]{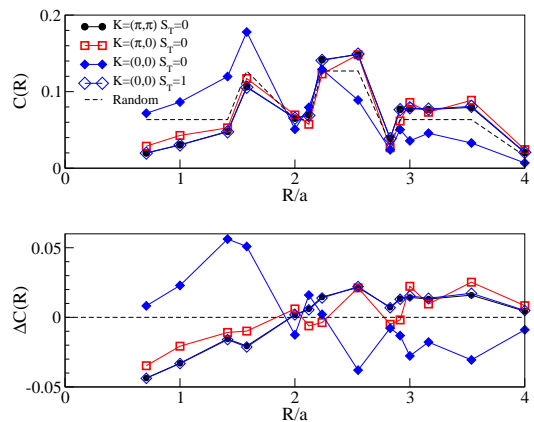}
\caption{\label{fig:bpFig4} Top: Probability of charge separation,
  $\langle C(R)\rangle$ for the lowest state in the
  $C_S=\frac{1}{2}$ subspace. Bottom: The difference from a random
  distribution. See text for details.}
\end{figure}
The expectation values $C(R)=\langle \hat{C}(R)\rangle$ are
shown in Fig.~\ref{fig:bpFig4} at various high-symmetry
points. The lowest states with $K=(\pi,\pi)$ $S_T=0$ and $K=(0,0)$ $S_T=1$
have the same $C(R)$, confirming that they are indeed linked by
a $Q=(\pi,\pi)$ magnon excitation.

$\Delta C(R)$ features a monotonic increase with R for the GS,
a small peak of correlation at $R=2a$ for $(\pi,0)$, and local attraction for
the $S=0$ $K=(0,0)$ state. Section IV.A shows that the $\sim 2J_{dd}$ single-polaron bandwidth is the energy difference between the GS
with extended $\Delta C(R)$ and the $K=(0,0)$ $d_{x^2-y^2}$ state with local $\Delta C(R)$.
The low energy two-particle band is thus a result of the kinetic energy competing with an attractive
energy which induces a local $d_{x^2-y^2}$ pair.

\subsubsection{Polaronic Nature}

In the single-hole case,\cite{spinpolaron} at low-energies
the mobile carrier is well described as the spin-$\frac{1}{2}$ 3SP's,
$|\Uparrow\rangle$ and $|\Downarrow\rangle$ in Table \ref{tab:2h_3sp}, which are correlated states of the oxygen spin with its two neighboring copper
spins. For the exact ground state
 $\langle H_{J_{pd}}\rangle\sim -0.9J_{pd}$, close to the $-J_{pd}$
energy of
the exact 3SP. The two-hole
solutions yields $\langle H_{J_{pd}}\rangle\sim -1.8J_{pd}$,
showing that thinking in terms of 3SP  is still valid and fruitful.
%Determining the spin-spin correlation between the
%two 3SP requires phase information among the two oxygen and the
%four copper spins involved. This is not a straightforward exercise because
%annihilating copper spins would take the wavefunction outside of the
%singly-occupied states kept in
%the numerical basis.
%It is instructive to first consider the eight single-hole eigenstates of
%$H_{J_{pd}}$ listed in Tab.~\ref{tab:2h_3sp}.
In this scenario, the oxygen holes could neighbor the same copper spin to form a
5-spin object, but $C(R)$ of Fig.~\ref{fig:bpFig4}
shows that the probability for this is low. Ignoring these shared-copper
configurations, the wavefunction contains two polarons involving a
total of six spins. Noting that $\langle H_{J_{pd}}\rangle\sim
-1.8J_{pd}$, the single-polaron levels in Tab.~\ref{tab:2h_3sp}
suggest that
the dominant part of the wavefunction contains four possible 3SP pairs
with $\frac{\langle H_{J_{pd}}\rangle}{J_{pd}}=-2$:
$|\Uparrow\rangle|\Downarrow\rangle$,$|\Uparrow\rangle|
\Downarrow\rangle$,$|\Uparrow\rangle|\Uparrow\rangle$,
and $|\Downarrow\rangle|\Downarrow\rangle$. Other pair configurations
would have
$\frac{\langle
  H_{J_{pd}}\rangle}{J_{pd}}\in\left\{0,\pm\frac{1}{2},\pm
1\right\}$, and can be ignored. This can be achieved using a 3SP-pair projector
\begin{equation}
P_{3SP}=\prod_{\Delta\lambda\in\{0,\pm\frac{1}{2},\pm 1\}}\frac{\widehat{H}_{J_{pd}}/J_{pd}-\Delta\lambda}{-2-\Delta\lambda},
\end{equation}
wherein terms in the product vanish for the eigenvalues of the
excluded levels, while scaling the relevant $\langle
  H_{J_{pd}}\rangle=-2J_{pd}$ levels to unity.
$P_{3SP}$ thus projects out the subspace of having two 3SP's
  with any spin-spin correlation. All low energy states have $\gtrsim 0.78$ probability in the $P_{3SP}$-projected subspace, roughly the square of single-hole
  solution's $\sim 0.9$ overlap with a 3SP.\cite{spinpolaron} It is
  thus clear that the extra holes form three-spin polarons.

Section IV.A shows that the low-energy two-hole states are total spin
$S=0$ singlets. It is thus interesting to check if the wavefunctions can be described simply
by a $S=0$ AFM background plus a 3SP singlet pair,
$\frac{1}{\sqrt{2}}\left(|\Uparrow\rangle|\Downarrow\rangle-
|\Uparrow\rangle|\Downarrow\rangle\right)$.
In Appendix B, we derive a numerical operator
\begin{equation}
S_{3SP}(R)=\sum_{|\eta|=R}s_{3SP}(\eta),
\end{equation}
which measures the probability of having
such a singlet as a function of charge-charge separation $R$, within
the subspace projected by $P_{3SP}$.
The value is normalized with the analogous triplet measure, defined in a
similar fashion:
$$\sum_R \left(S_{3SP}(R)+T_{3SP}(R)\right)=1.$$

The singlet correlation should be compared to $\sum_R P'(R)=1$, the random distribution
of two 3SPs spread over the 64 oxygen sites but with no shared copper spin.
The probability of singlet correlation is $\frac{1}{4}$ in a paramagnetic state.
The difference of interest is then
\begin{equation}
\Delta S_{3SP}(R)=S_{3SP}(R)-\frac{P'(R)}{4}.
\end{equation}

\begin{figure}[t]
\centering
\includegraphics[width=0.8\columnwidth]{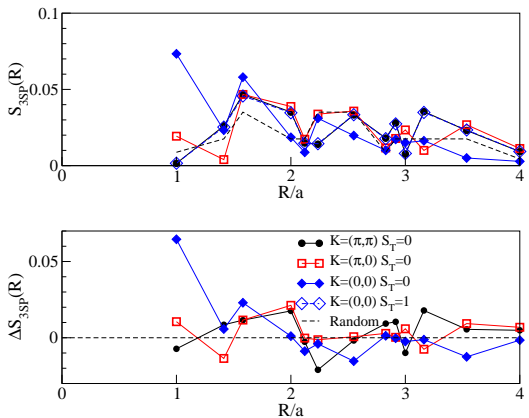}
\caption{\label{fig:bpc3spsinglet} Top:  Probability of singlet 3SP
  pair, $\langle S_{3SP}(R)\rangle$ for the lowest states in
  the $C_S=\frac{1}{2}$ subspace. Bottom: The difference from a
  randomly distributed paramagnetic configuration.}
\end{figure}

Figure \ref{fig:bpc3spsinglet} shows $S_{3SP}(R)$ and $\Delta
S_{3SP}(R)$ for the high-symmetry points eigenstates. It is evident
that all states have enhanced short-range singlet nature as compared
to the random distribution -- a feature absent from previous studies
using small clusters or the t-t'-J model which does not distinguish
oxygen and copper sites.  This short-range nature is not an artifact
of small cluster squeezing the two polarons together because the
cluster does allow the two holes to separate further apart, as confirmed
by the long-range nature of charge correlation in
Fig.~\ref{fig:bpFig4}.

$\Delta S_{3SP}(R)$ tells only part of the story because
it is counter intuitive to have the two polarons with $R$-dependent singlet-triplet
ratio as indicated by Fig.~\ref{fig:bpc3spsinglet}. The total spin is
$S_T=0$ so the $N-4$ copper spins not neighbored by the oxygen charges
must carry a spin of 1 when the two polarons tend to form a
triplet. The 6-spin description, which is sufficient to describe the
location of charge, is not sufficient to capture the spin carried by
the polarons. The polarons produce a spin disturbance with a
$R$-dependent spatial extent beyond the two copper spins sandwiching the
oxygen charge. Note that the non-local spin disturbance around a polaron
is already observed in the single-hole case.\cite{spinpolaron}

\section{Discussion}

%\subsection{Short-range attraction}

Section IV shows that in the GS, the two holes are not correlated spatially, since the probability $\Delta C(R)$ to find them at $R\le 2a$ is less than for randomly distributed holes. This suggests that if the binding energy indeed remains negative in the limit $N\rightarrow\infty$, the pair would have to be correlated in k-space. One needs to be able to study larger clusters in order to fully settle this issue. However, local pairing correlations are seen for the two holes in the S=0, K=0 state, showing that the competition between kinetic energy and a local attractive potential. For an energy cost of a single polaron's bandwidth, $\sim 2J_{dd}$, the state changes from the $K=(\pi,\pi)$ delocalized GS to a $K=0$ $d_{x^2-y^2}$ local pair. This strong momentum dependency suggests that, amidst the presence of a local $d_{x^2-y^2}$ attraction, a local-pairing scenario cannot completely explain the full problem, at least within this model in the $n=2/32$ low-doping scenario.

Although we have broken the technological limit, two holes in a 32-CuO$_2$-unit-cell finite lattice cannot rigorously model all aspects of an infinite system. For example, the possibility of a quantum critical point in the 4.3\% low-doping regime\cite{qcp} cannot be tested. There are other questions: what is the implication of the $\sim 2J_{dd}$ penalty when long-range AFM order due to $J_{dd}$ is destroyed? Is there a better measure of the binding energy than referencing from a single polaron along with a AFM background? What are the translational and point-group symmetries of the overall wavefunction of more than two holes? Finite-size scaling beyond N=32 would certainly provide a more robust description of the physics but would require more advanced approaches and better technologies.

Next we point out that the weak local attraction, if it ever gains prominence as doping is increased, would not lead to the unrealistic real-space clustering of holes as doping concentration is raised. The only hint of clustering mechanism is the attraction for local $d_{x^2-y^2}$ correlated holes as in the $K,S_T=0$ state. The nature of this attraction can be grasped from the $R\le a$ local peaks of singlet tendency in Fig.~\ref{fig:bpc3spsinglet}.
By expanding the 6-spin wavefunctions
$\frac{1}{\sqrt{2}}\left(|\Uparrow\rangle|\Downarrow\rangle\pm|\Downarrow
\rangle|\Uparrow\rangle\right)$ and taking the
$\overline{S}\cdot\overline{S}$ between a
copper spin from the first 3SP and one from the second 3SP, the
expectation value is $-\frac{5}{18}J_{dd}$ if they are singlet and
$+\frac{1}{18}J_{dd}$ if they are triplet. Two 3SPs can
take advantage of this energy lowering by forming a singlet, but only if they are
separated by an empty oxygen site which mediates an ordinary
$\overline{S}\cdot\overline{S}$ Heisenberg AFM bond. The two-hole
numerical solution indeed shows a local maximum when $R/a\leq 2$
(Fig.~\ref{fig:bpc3spsinglet}). Now, by expanding the 12-spin wavefunctions obtained as the direct product of two 3SP singlet pairs:
$\frac{1}{2}\left(|\Uparrow\rangle|\Downarrow\rangle-|\Downarrow\rangle|\Uparrow
\rangle\right)\bigotimes\left(|\Uparrow\rangle|\Downarrow\rangle-|\Downarrow\rangle|
\Uparrow\rangle\right)$,
the expectation value of $\overline{S}\cdot\overline{S}$ between a
copper spin from the first pair and one from the second pair is
exactly zero. Therefore, more than two holes cannot lower their energy by clustering.

Lastly, we point out features of the two-hole solution that is expected to prevail as the cluster size $N$ is increased. The doped charge would form a core of $3SP$, which is surrounded by spin disturbances which determine the spin of the polaron, in agreement of the one-hole scenario\cite{spinpolaron}. The extent of spin disturbance varies with hole-hole separation. The R-dependent polaron-polaron correlations would not change drastically because the $N=32$ cluster can accommodate the two holes without artificially forcing them together as in the $N=16$ case. Due to this fact, the observed cross-over between locally and non-locally correlated states within a $\sim 2J_{dd}$ energy window is expected to be robust.

\section{Conclusions}
To summarize, we have extended our previous work
\cite{spinpolaron,octapartite} to study two-hole states in CuO$_2$
planes, in the context of the spin-polaron model of
Ref.~\onlinecite{spinpolaron}. Our numerical approach bypassed various
technical limitations and extracted the explicit low-energy
wavefunctions of two holes injected into a half-filled system with 32
copper and 64 oxygen sites.

The $N=32$ solution was found to be different from the $N=16$
solution. Similar to the $N=32$ two-hole t-t'-J model, the GS was
found at $K=(\pi,\pi)$, with $\Delta E\sim\frac{1}{N}$ $Q=(\pi,\pi)$ magnon
excitation. The binding energy was found to be $\left(-0.091\pm
0.272\right)J_{dd}$. In contrast to the $t-t'-J$ model, we found
the lowest state at $K=(0,0)$, without a
$Q=(\pi,\pi)$ magnon, to have be a $d_{x^2-y^2}$
locally bound state.

Further analysis of the wavefunctions revealed that the charge
carriers are 3SPs, $|\Uparrow\rangle$ and $|\Downarrow\rangle$, even in
this multi-hole scenario, but the polarons' spin disturbances are
extended in range, involving more than the two copper spins
sandwiching the 3SP.  From the correlation values, we established that
the low-energy band results from the competition between kinetic
energy and a local attractive potential which induces $d_{x^2-y^2}$
pair.

Lastly, we showed that real-space hole clustering is unlikely at higher concentration and also noted aspects that are expected to be robust for larger systems size. Study of higher-doping scenario could be interesting but would require more advanced technologies.

%We pointed out aspects that are expected to be robust for large
%systems at higher doping. We then speculated on a  possible
%doping-dependent mechanism for short-range singlet attraction: 3SPs
%are formed, destroying AFM order in their vicinity. There is no
%evidence of clustering mechanism beyond two particle pairs.  As AFM
%correlations increase and the kinetic energy rises, short-range 3SP
%singlet pairs are formed as individual bipolaron entities which tend
%to be more than one empty oxygen sites away from other pairs. The
%mechanism is destroyed at higher doping when the available space does
%not allow bipolarons to stay that far apart. The hard limit is at
%$25\%$ doping, when bipolarons can no longer be more than one empty
%oxygen site apart. This scenario provides an incentive for studying larger systems
%with more powerful techniques.

\acknowledgements We thank R. Thomale for discussions, P. W. Leung for
sharing two-hole data for the t-t'-J model, I. Elfimov and Westgrid
for tech support, and CFI, CIfAR, CRC and NSERC for funding.

\appendix

\section{Implementation of the octapartite truncation for the AFM background}

First, we define singlet and triplet creation operators for the two
oxygen holes, when located at sites $l\ne l'$:
\begin{eqnarray}
s^\dag_{l,l'}&=&\frac{1}{\sqrt{2}}(p^\dag_{l\uparrow} p^\dag_{l'\downarrow}-p^\dag_{l\downarrow} p^\dag_{l'\uparrow})\nonumber\\
t^\dag_{-1,l,l'}&=&p^\dag_{l\downarrow} p^\dag_{l'\downarrow}\nonumber\\
t^\dag_{0,l,l'}&=&\frac{1}{\sqrt{2}}(p^\dag_{l\uparrow} p^\dag_{l'\downarrow}+p^\dag_{l\downarrow} p^\dag_{l'\uparrow})\nonumber\\
t^\dag_{1,l,l'}&=&p^\dag_{l\uparrow} p^\dag_{l'\uparrow}.\label{eq:2h_singlettriplet}
\end{eqnarray}

Recall that $S$ is the total spin of the $N$ Cu spins (the AFM
background).
According to quantum mechanical angular momentum addition, upon
introducing the two doping holes, the $N+2$ spins can add up to a total
spin $S_T$ by mixing a two-hole singlet with a background of total
spin $S=S_T$, or by mixing
a two-hole triplet with a background of total spin $S=S_T, S_T\pm 1$. Taking $S_T^z=0$,
any spin background can be specified orthonormally with respect to a position $l$:
\begin{equation}
\square^\dag_{l,l'}|\alpha\rangle_l\equiv\Bigg\{\begin{matrix}
s^\dag_{l,l'}|\alpha,S^z=0\rangle_l\\
\sum_{z=-1}^1 c(z,S,S_T)t^\dag_{z,l,l'}|\alpha,S^z=-z\rangle_l
\end{matrix}\label{eq:2h_singlettripletbackground}
\end{equation}
Here,
$\alpha$ denotes a particular group of $2S+1$ spin configurations related by
the total spin raising and lowering operators $S^{\pm}=\sum_l S^{\pm}_l$, summed over all Cu sites.
The total spin of this $\alpha$ group can be $S=S_T$ for two-hole singlet and $S=S_T,S_T\pm 1$
for two-hole triplets. Due to the choice of the overall projection $S_T^z=0$, the two-hole singlet would
mix only with backgrounds with $S^z=0$. The two-hole triplets would mix with three
different projections $|\alpha,S^z=-1\rangle_l$, $|\alpha,S^z=0\rangle_l$, and
$|\alpha,S^z=1\rangle_l$ from the group $\alpha$. The weight $c(z,S,S_T)$ is the
Clebsch-Gordon coefficients for mixing these three states with the three two-hole
triplets to achieve a state with total spin $S_T$ and $S_T^z=0$.

Exploitation of the translational symmetry is performed by the use of a Fourier series of
the form
\begin{equation}
\sim\sum e^{iKl}\square^\dag_{l+\epsilon,l'+\epsilon'}|\alpha\rangle_l;\nonumber
\end{equation}
however, care must be taken to ensure orthonormality. Due to the commutation relation of
the triplet and singlet (Eq.~\ref{eq:2h_singlettriplet}), the Fourier series is not straightforward
for $\epsilon=\epsilon'=\epsilon_{x/y}$ when both oxygen holes occupy x- or y-rung oxygen orbitals
(see Fig. \ref{fig:cuo2}a). The specification of these two-hole configurations requires
the two orthogonal periodic lattice vector $L_0$ and $L_1$ of length $\sqrt{N}$ for the 2D,
N-unit-cell lattice. Defining
\begin{equation}
\delta l=(l'_0-l_0,l'_1-l_1),
\end{equation}
most hole-hole configurations can be classified in the region
\begin{eqnarray}
0\leq\delta l_0<\frac{L_0}{2}&,&0<\delta l_1<\frac{L_1}{2}\nonumber\\
0<\delta l_0<\frac{L_0}{2}   &,&\delta l_1=\frac{L_1}{2}\nonumber\\
0<\delta l_0<\frac{L_0}{2}   &,&-\frac{L_1}{2}<\delta l_1\leq 0\nonumber\\
\delta l_0=\frac{L_0}{2}   &,&-\frac{L_1}{2}<\delta l_1< 0.
\end{eqnarray}
These states can be expressed using $N$-term Fourier series
\begin{equation}
|\square_{xx/yy},\delta l,\alpha,K\rangle=\frac{1}{\sqrt{N}}\sum
e^{iKl}\square^\dag_{l+\epsilon_{x/y},l+\delta l
  +\epsilon_{x/y}}|\alpha\rangle_l
\end{equation}
Because the two oxygen holes are indistinguishable fermions
(Eq.~\ref{eq:2h_singlettriplet}), there are three remaining
$\delta l$ values which require special attention due to the
periodic boundary condition.
\begin{eqnarray}
\delta l_0=\frac{L_0}{2} &,& \delta l_1=0\nonumber\\
\delta l_0=0 &,& \delta l_1=\frac{L_1}{2}\nonumber\\
\delta l_0=\frac{L_0}{2} &,& \delta l_1=\frac{L_1}{2}
\end{eqnarray}
The number of terms in the Fourier series depends on the spin background
translated by $\delta l$: $T_{\delta l}|\alpha\rangle_l$. If such a
translation yields an orthogonal state, $_l\langle\alpha|T_{\delta l}|\alpha\rangle_l=0$,
the Fourier series still has $N$ terms. For the example
of $\delta l=(\frac{L_0}{2},0)$,
\begin{widetext}
\begin{eqnarray}
|\square_{xx/yy},\delta l,\alpha,K\rangle
&=&\frac{1}{\sqrt{N}}\sum e^{iKl}\square^\dag_{l+\epsilon_{x/y},l+\delta l +\epsilon_{x/y}}|\alpha\rangle_l\nonumber\\
&=&\frac{1}{\sqrt{N}}\sum_{l_1=0}^{L_1-1}e^{iK_1l_1}\sum_{l_0=0}^{\frac{L_0}{2}-1}e^{iK_0l_0}\square^\dag_{l+\epsilon_{x/y},l+\delta
  l
  +\epsilon_{x/y}}\left(1+s_{\square}e^{iK_0\frac{L_0}{2}}T_{\frac{L_0}{2}}\right)|\alpha\rangle_l\label{eq:2h_xxyy},
\end{eqnarray}
\end{widetext}
where $s_{\square}$ is the sign change due to hole-swapping in the
singlet or triplet $(\square^\dag_{a,b}=s_{\square}\square^\dag_{b,a})$.
For the case of $\langle\alpha|T_{\delta l}|\alpha\rangle_l=\pm 1$, the above
expansion makes clear that there can be only $\frac{N}{2}$ terms in
the Fourier series due to the term $\left(1+s_{\square}e^{iK_0\frac{L_0}{2}}T_{\frac{L_0}{2}}\right)$.
For this case the series has the form
\begin{widetext}
\begin{equation}
|\square_{xx/yy},\delta l,\alpha,K\rangle
=\sqrt{\frac{2}{N}}\sum_{l_1=0}^{L_1-1}e^{iK_1l_1}\sum_{l_0=0}^{\frac{L_0}{2}-1}e^{iK_0l_0}\square^\dag_{l+\epsilon_{x/y},l+\delta
  l +\epsilon_{x/y}}|\alpha\rangle_l.\label{eq:2h_xxyy_b}
\end{equation}
\end{widetext}
The formulation for the case where one hole occupies $p_{l+\epsilon_x}$ and
the other  $p_{l+\epsilon_y}$ is straightforward. All values of $\delta l$
are unique and the Fourier series has the form
\begin{equation}
|\square_{xy},\delta l,\alpha,K\rangle=\frac{1}{\sqrt{N}}\sum
e^{iKl}\square^\dag_{l+\epsilon_{x},l+\delta l
  +\epsilon_{y}}|\alpha\rangle_l.\label{eq:2h_xy}
\end{equation}

Therefore, a full orthonormal Hilbert space can be specified by the
states $|\square_{xy},\delta l,\alpha,K\rangle$ and
$|\square_{xx/yy},\delta l,\alpha,K\rangle$.
Translational symmetry is specified by the quantum number K. Total
spin and its projection are specified by the singlet/triplet nature of
the two oxygen holes ($\square$) (singlet-triplet) in conjunction to
the spin background $\alpha$.

The above formulation allows the enumeration of different
oxygen-oxygen configurations for the two doping holes, for a given $|\alpha\rangle$
background. An enumeration of the different $|\alpha\rangle$ states
is now the missing step for computation. This is not a trivial
exercise because the computation for large $N$ requires that a large
number of possible
$|\alpha\rangle$ states be discarded according to the
truncation criterion of Eq.~\ref{eq:truncation}. Note that this criterion is
based on the total spin of each sublattice, and the basis $\alpha$ is
built according to quantum mechanical spin addition.

An arbitrary enumeration of $|\alpha\rangle$ basis would not yield an
efficient computation because of several challenges. First, the
truncation of less important states should be controllable
systematically and flexible enough to adapt to the many unknowns of
doped systems. Second, within the truncated basis transformed away
from the natural $z$-projected description, there should be a fast way
of indexing initial and final states upon Hamiltonian
operations. Third, one should have apriori knowledge about the
identically-zero matrix elements in order to avoid them for sparse
matrices' performance scaling.

The above formulation specifies a basis state in terms of a Fourier
sum over a particular reference site $l$. The spin background
$|\alpha\rangle_l$ and one oxygen hole $p^\dag_{l+\epsilon,\sigma}$
are referenced at $l$ while the other oxygen hole is located at
$\delta l$ away from the reference oxygen hole. The hopping terms
$T_{pp}$, $T_{swap}$ and $J_{pp}$ operate on both oxygen
holes. Hopping of the non-reference oxygen hole would not change the
reference point $l$; however, hopping of the oxygen
hole at ${l+\epsilon}$ would yield a state with no oxygen at
$l+\epsilon$. This simply means that the final state can be specified
at a new referencing position $l'$ with a phase shift of $e^{i K
  (l-l')}$, and that the new spin background is obtained by a
translation
$|\alpha'\rangle_{l'}=\widehat{Tr}(l-l')|\alpha\rangle_{l}$. It is
thus advantageous to build the $|\alpha\rangle_l$ spin basis such that
the
construction is the same upon translation.

\begin{figure}[t]
\centering
\includegraphics[width=\columnwidth]{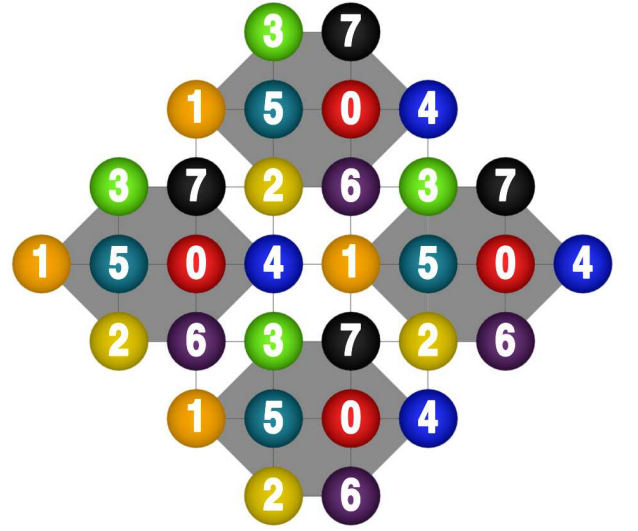}
\caption{\label{fig:octa} $N=32$ cluster divided into eight
  groups labbeled $0=7$. The spins within each group are connected by multiples of
  $(2a,\pm 2a)$. Spins from a particular group are always in the same
  environment, {\em eg} 1 always 4,5,6 and 7 as neighbors.}
\end{figure}

The octapartite approach introduced in Ref. \onlinecite{octapartite} satisfies all of
the above requirements. First, the copper spins are divided into
eight groups as shown in Fig.~\ref{fig:octa}. For each group of
$\frac{N}{8}$ spins, we start with the $z$-projected representation of
$2^{\frac{N}{8}}$ states and transform into a Clebsch-Gordan basis
$|s_{\frac{N}{8}},s^z_{\frac{N}{8}}\rangle$.  Each $z$-projected basis
state is represented using bits of an unsigned integer and each
Clebsch-Gordan series is stored as a sparse vector. With a particular
enumeration of $|s_{\frac{N}{8}},s^z_{\frac{N}{8}}\rangle$ states, we
mix two identical enumerations to build
$|s_{\frac{N}{4}},s^z_{\frac{N}{4}}\rangle$ according to
Clebsch-Gordan addition. Then with two identical enumerations of
$|s_{\frac{N}{4}},s^z_{\frac{N}{4}}\rangle$, we build a single
enumeration of $|s_{\frac{N}{2}},s^z_{\frac{N}{2}}\rangle$. Finally,
we can similarly enumerate the overall $|\alpha\rangle$ background as
$|s_{N},s^z_{N}\rangle$ from the enumeration of
$|s_{\frac{N}{2}},s^z_{\frac{N}{2}}\rangle$.  This is a recursive
enumeration procedure. At each stage, the states are indexed by a
non-negative integer, \texttt{state\_index}. The state represented by
a particular index value can be derived from the ``parent" basis with
half the number of spins using, for example, the following loop
structure:

\begin{widetext}
\begin{lstlisting}
size_t state_index=0;
//in general total spin can be odd multiples of 1/2, so I worked with 2S
const int min2S=minimum_2_times_total_spin;
const int max2S=maximum_2_times_total_spin;
const int minSub2S=minimum_2_times_sub_lattice_spin;
const int maxSub2S=maximum_2_times_sub_lattice_spin;

for(int ss=min2S;ss<=max2S;ss+=2){//states with higher total spin have higher indices
  for(int ssa=0;ssa<=maxSub2S;ssa+=2){//loop over the left ket
    for(int ssb=0;ssb<=maxSub2S;ssb+=2){//loop over the right ket
      if(Clebsch_Gordan_coefficients==0) continue;

      const size_t na=number_of_(ssa+1)_blocks_with_total_spin_ssa;
      const size_t nb=number_of_(ssb+1)_blocks_with_total_spin_ssb;
      for(size_t aa=0;aa<na;++aa){
        for(size_t bb=0;bb<nb;++bb){
          for(int zz=-ss;zz<=ss;zz+=2,++state_index){//loop over the ss+1 components of z-projection
            fprintf(stderr, "State %llu has s=%4f and sz=%+4f, product of",
                    state_index,0.5*ss,0.5*zz);
            fprintf(stderr, "block %llu/%llu of the s=%4f sector of left ket and ",
                    aa,na,0.5*ssa);
            fprintf(stderr, "block %llu/%llu of the s=%4f sector of right ket",
                    bb,nb,0.5*ssb);
          }//zz
        }//bb
      }//aa
    }//ssb
  }//ssa
}//ss
\end{lstlisting}
\end{widetext}

The benefit of this enumeration is an instantaneous ``reverse lookup".
When performing Hamiltonian operations, one is really interested
in the non-zero overlap between the outgoing states and those in the
orthonormal basis. The naive way is to compute the dot product against
all basis states, but this adds a $O(n)$ layer on top everything else
and is detrimental in the case of large systems. Under the above
looping scheme, an increasing \texttt{state\_index} is associated with
increasing values of \texttt{ssa}, \texttt{aa}, \texttt{ssb}, and
\texttt{bb}. Because the index shift for states of these combinations is known a priorily, any arbitrary ket
$|\sigma_a,\sigma^z_a\rangle|\sigma_b,\sigma^z_b\rangle$, is trivially
associated with these four indices so states with non-zero overlap are
known immediately, with an $O(1)$ reverse lookup operation. Non-zero
matrix elements can thus be computed efficiently by changing
\texttt{ssa}, \texttt{aa}, \texttt{ssb}, and \texttt{bb} before the
reverse lookup. The neighboring pattern in Fig.~\ref{fig:octa} eases
the determination of spin backgrounds upon translation. The
determination of matrix elements is trivially parallelizable. It is
apparent that truncation according to the criterion of
Eq.~\ref{eq:truncation} can be performed at last stage of mixing by
simply tuning the value of \texttt{minSub2S}.

\section{Singlet correlator between two three-spin polarons}

If we consider only the 6 spins involved in the two 3SPs, the projected wavefunction is a superposition of the four
possible 3SP pairs:
\begin{eqnarray}
|\phi_6\rangle&=&a\frac{|\Uparrow\rangle|\Downarrow\rangle-|\Uparrow\rangle|\Downarrow\rangle}{\sqrt{2}}+b|\Uparrow\rangle|\Uparrow\rangle\nonumber\\
&&+c\frac{|\Uparrow\rangle|\Downarrow\rangle+|\Uparrow\rangle|\Downarrow\rangle}{\sqrt{2}}
+d|\Downarrow\rangle|\Downarrow\rangle.\nonumber
\end{eqnarray}
Defining the oxygen-oxygen singlet operator for the two oxygen spins, $p^\dag_{1,\sigma}$ and $p^\dag_{2,\sigma}$
\begin{equation}
s^\dag_{1,2}=\frac{1}{\sqrt{2}}\left(p_{1,\uparrow}^\dag p_{2,\downarrow}^\dag-p_{1,\downarrow}^\dag p_{2,\uparrow}^\dag\right),
\end{equation}
the probability of finding a oxygen-oxygen singlet in $|\phi_6\rangle$ is
\begin{equation}
\left\langle s^\dag_{1,2}s_{1,2}\right\rangle =\frac{3}{9}a^2+\frac{2}{9}(b^2+c^2+d^2).
\end{equation}

Solving for $a^2$ and
generalizing the expression to different hole-hole separations $\eta$,
the  singlet nature of the 3SP pair when at distance $\eta$ apart is
gauged by
%%%%%%%%%%%%%%%%%%%%%%%%%%%%%% EQUATION %%%%%%%%%%%%%%%%%%%%%%%%%%%%%%
\begin{equation}
s_{3SP}(\eta)=\frac{\langle
  P_{3SP}\left(9\hat{s}(\eta)-2\hat{c}(\eta)\right)P_{3SP}\rangle}{\langle
  P_{3SP}\sum_{\eta'}\hat{c}(\eta')P_{3SP}\rangle}
\end{equation}
%%%%%%%%%%%%%%%%%%%%%%%%%%%%%%%%%%%%%%%%%%%%%%%%%%%%%%%%%%%%%%%%%%%%%%
where
$$ \hat{s}(\eta)=\sum_{l,\epsilon}
s^\dag_{l+\epsilon,l+\epsilon+\eta}s_{l+\epsilon,l+\epsilon+\eta}
$$ measures the probability of the two O holes to be in a singlet at
distance $\eta$ apart, irrespective of what the Cu spins are
doing. Thus, $s_{3SP}(\eta)$ is the probability, within the projected
subspace, of the two oxygen holes and the AFM background to cooperate
to form a 3SP singlet pair separated by a distance $\eta$. The value
ranges from zero for no singlet nature to unity for pure singlet at
this hole-hole separation; however, these two extreme values are not
possible due to, for example, the spatial spreading required to lower
the kinetic energy of $T_{pp}$ and $T_{swap}$. The measure can be
summed as a function of hole-hole separation
\begin{equation}
S_{3SP}(R)=\sum_{|\eta|=R}s_{3SP}(\eta),
\end{equation}
which is normalized with the analogous triplet measure, defined in a
similar fashion:
$$\sum_R \left(S_{3SP}(R)+T_{3SP}(R)\right)=1.$$

The singlet correlation should be compared to $\sum_R P'(R)=1$, the random distribution
of two 3SPs spread over the 64 oxygen sites but with no shared copper spin.
The probability of singlet correlation is $\frac{1}{4}$ in a paramagnetic state.
The difference of interest is then
\begin{equation}
\Delta S_{3SP}(R)=S_{3SP}(R)-\frac{P'(R)}{4}.
\end{equation}

% Create the reference section using BibTeX: \bibliography{basename of
%.bib file}

\end{document}